# Stress in Spin-Valve Nanopillars due to Spin Transfer

Hao Yu [a]

*Mathematics and Physics Centre, Xi'an Jiaotong-Liverpool University,Suzhou, 215123, China*

[Abstract] We report a mechanical effect in spin-valve nanopillars due to spin transfer. A polarized current carrying electron spins transfers torque to local magnetization and leads to a magnetic switching of free layer. Like classical Einstein-de Haas effect, the conservation of angular momentum needs the free layer to offset the change of angular momentum and then a mechanical rotation occurs. The free layer is not free standing, so the mechanical angular momentum will be revealed as stress and strain. We study the effect of a spin induced stress in a nanopillar device with in-plane magnetization. Our calculations show that the tress in as device is dependent on frequency and the ratio of length/thickness and about 1 MPa at GHz. It is concluded that the stress owing to spin transfer is much less than the internal stress of film and does not introduce damage to the device.

Keywords: Einstein-de Haas effect, spin transfer torque, stress

______________

[a]Email: hao.yu@xjtlu.edu.cn

Electrons contain two degrees of freedom, charge and spin. Spin has been proved to be angular momentum since 1906 when Einstein and de Haas found that a magnetic rod can be rotated applied a magnetic field [1]. The conservation of angular momentum leads to many classical and quantum effect. For example, spin transfer torque effect, found by Slonczewski [2], is due to the angular momentum transfer when a polarized current traverses a spin valve with nano-scale contact or a magnetic domain wall [3].

In recent years, spin transfer torque effect has been intensively studied because of its potential application in data storage and logic device technology [3]. To use current instead of magnetic field to drive domain-wall motion or alter the orientation of magnetization is an innovation, which brings higher density data storage and faster read/write speed. Spin transfer torque effect is also investigated in the field of NEMS because of the angular moment transfer process [4-7]. It has been demonstrated in both theory and experiment that a spin-polarized current can induce a mechanical rotation at magnetic/non-magnetic interface of a suspended nanowire[8], which is like the classical Einstein-de Haas effect. And a similar effect, spin-driven mechanical rotation, also occurs at a nanowire with domain wall [7].

Then how this spin-mechanics coupling phenomenon works at a nanopillar spin valve device? Does this effect introduce damage to the device due to the mechanical deformation by the transfer of angular momentum from electrons to crystal lattice? Though a fixed device cannot rotate, a potential mechanical angular momentum is to exert stress and stain at the nanometer scale structure and may influence the performance of it especially in the case of high frequency.

In this work, a nanopillar structure device is considered. The device has 3 parts, as shown in Fig.1, including a thick magnetic pinned layer (PL), a non-magnetic spacer layer(SL) and a thin magnetic free layer(FL). A current traverses PL vertically and is polarized. When the polarized current flows to the free layer and if the current density is strong enough (above a threshold value), the magnetic orientation of the FL will be switched because of the interaction between the spin of flowing electrons and local electrons. The spin transfer process occurs and some of the angular momentum of the current is transferred to the local magnetic moment. The spin resource from conductive current can be considered as an effective magnetic field and in which, for the FL, angular momentum must be conserved, because there is no external torque applied on FL. The magnetization of FL has been changed and there must be another angular momentum to balance this change. The spin transfer to lattice should be considered. From the Einstein-de Haas effect, we know that, a rod can rotate when it is exerted by an external magnetic field. The conservation of angular momentum makes the rod should vary its rotation angular momentum to offset the change of the magnetization. Like the Einstein-de Haas effect, a mechanical angular momentum occurs at FL to obey the conservation law. However, the sandwich films are deposited on substrate, so the FL can not be rotated and then the lattice rotation will lead to stress and strain in the FL. Moreover, the thickness of spacer layer is less than the diffusing length of spin, so it is supposed that the transfer to mechanical rotation doesn't take place at the magnetic/non-magnetic interface between PL and SL.

The geometry of this device: the cross section of the pillar is rectangle with L=100 nm length and a=50 nm width. The thickness of the pinned layer, which is not important in our calculation, is much thicker than the FL. The FL is with b=5nm thickness.

The pinned layer and free layer are both in-plane magnetized. And it is supposed that the magnetization direction is parallel to the length direction of FL. The FL switches between two saturation magnetization direction (parallel or antiparallel to the magnetization of PL). On the basis of the law of the conservation of angular momentum, the total angular momentum of FL must remain unvaried, and magnetization FL therefore acquires an angular impulse that is inverse with respect to the axis of magnetization. The mechanical torque is equal to the time rate of change of angular momentum[9], as

$$T = 1/\gamma \cdot \Delta M \cdot f, \qquad (1)$$

where T is the mechanical torque, $\gamma$ is the gyromagnetic ratio and $\Delta M$ is the change of magnetic moment of FL, and $f$ is the driving frequency of polarized current. $\gamma = eg_e/(2m_e)$, where $m_e$ is the rest mass of electron, $e$ the electron charge and $g_e \approx 2$ is g-factor. For the FL switches between two antiparallel states, $\Delta M$ can be calculated by

$$\Delta M = 2 M_s \cdot Vol. \qquad (2)$$

where $M_s$=1000emu/cm$^3$=10$^6$A/m (from experiment as ref [10].) and $Vol$=L×a×b, is the volume of FL.

Then we obtain the torque T=2.84×10$^{-28}f$ N·m, which is propotional to the frequency $f$ of pulse current. The torque applied FL generates a torsional deformation [11, 12]. The FL can be treated as a rectangle plate with length L= 100nm, width a=50nm and thickness b= 5nm. Then the maximum shearing stress [13] occurs along the center line of the wider face of FL is equal to

$\tau_{max}=T/(c_1 \cdot ab^2)$, (3)

where $c_1$ is a coefficient and dependent on the ratio a/b, and if a/b=10, the value of $c_1$ is 0.312[13], as shown in Fig. 2.

The evaluated $\tau_{max}$ for our model is $7.29 \times 10^{-4} f$ Pa. In the case of low frequency, the stress is very small. However, in the high-frequency case, for example, where the current pulse is nanosecond (f is GHz), the stress can be nearly 1 *MPa*. The stress may also be negligible compared to the internal stress of a film, which is typically more than 100 Mpa.

By putting Eqation (1),(2) and (3) together, we can find that $\tau_{max}$ is actually propotional to L/b, the ratio of the length to the thickness of FL, which means, for fixed frequency, thinner the FL is, and larger the stress is.

Because the FL is not free standing, the resonance between the current frequency and the inherent frequency of the device can be neglected. The inherent frequency of the whole device is much less than the driven frequency of the current.

To be concluded, this calculation work investigates a potential mechanical effect induced by spin transfer. The law of conservation of angular momentum makes the free layer provide a rotational lattice deformation to offset the change of magnetization. The result shows that the stress is propotional to the frequency of the polarized current, and the ratio of the length to the thickness of free layer (if the magnetization is along the length direction). For a nanosecond-pulse current, the stress can be with the order of magnitude of 1MPa, which is much less than membrane stress and will not induce damage to the device.

We thank Shuai Dong for stimulating discuss. This work has been supported by National Natural Science Foundation of China (No. 11047187) and Research Development Fund of Xi'an Jiaotong-Liverpool University.

Figures and Captions:

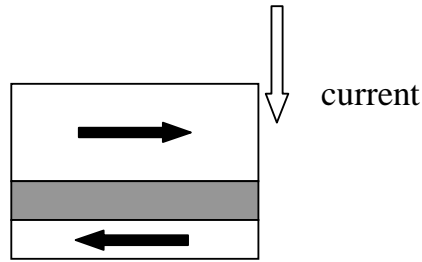

Fig. 1 A side view of a nanopillar spin valve. The sandwich structure composes of a pinned layer, a space layer(in gray) and a free layer. The current perpendicularly traverses the device. The magnetization direction of both pinned layer and free layer is in-plane.

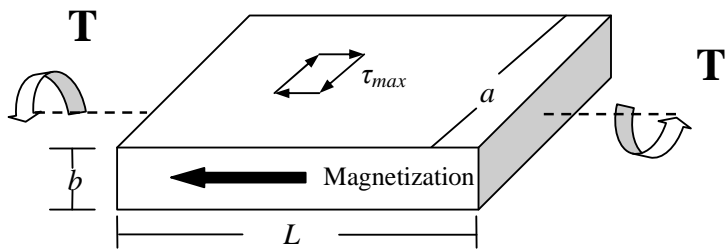

Fig. 2 The sketch of the stress of the free layer under a torque, which is in the direction parallel to the magnetization. The torsional mode figure is adapted from Ref. 13.